\documentclass[prb,twocolumn,showpacs,showkeys]{revtex4}

\usepackage{graphicx}
\usepackage{dcolumn}
\usepackage{bm}

\begin{document}


\title{Simulation of point defect diffusion in structures with local
elastic stresses}

\author{O. I. Velichko}

\email{oleg_velichko@lycos.com}

\affiliation{Department of Physics, Belarusian State University of
Informatics and Radioelectronics, 6, P.~Brovki Street, Minsk,
220013 Belarus}%


\begin{abstract}

The stress-mediated diffusion of nonequilibrium point defects from
the surface to the bulk of the semiconductor is investigated by
computer simulation. It is supposed that point defects are
generated in the surface region by ion implantation and during
diffusion pass over the local region of elastic stresses because
the average defect migration length is greater than the thickness
and depth of the strained layer. Within the strained layer point
defect segregation or heavily defect depletion occur if defect
drift under stresses is directed respectively in or out of the
layer. On the other hand, the calculations show that, in contrast
to the case of local defect sink, the local region of elastic
stresses practically does not change the distribution of defects
beyond this region if there is no generation/absorption of point
defects within the strained layer.

\end{abstract}


\maketitle

\section{Introduction}
In recent years decreasing dimensions of the integrated
components, \cite{Shao_03,Smith_05} especially with various
multilayered structures, \cite{Ghyselen_05} such as Si/SiGe,
\cite{Sawano_04,Buca_04,Leitz_06} and the use of different
nano-particles embedded in the crystalline matrix
\cite{Kanjilal_05,Coffin_06} have been the main trends of advanced
device technologies. Much attention is given to studying of the
defect kinetics and stress evolution during semiconductor
processing because defect and stress engineering can significantly
improve the device performance. \cite{Shao_03,Smith_05,Moroz_05}
For example, many efforts are directed to a study of the
stress-mediated diffusion of impurity atoms and point defects,
\cite{Park_95,Chaudhry_97,Daw_01,Aziz_01,Nazarov_04} including the
influence of elastic stresses on the drift of point defects near
the surface or at the interfaces,
\cite{Velichko_99,Gaiduk_03,Gaiduk_05} and near different
inhomogeneities of semiconductor crystals.
\cite{Park_95,Loiko_02}. Also, many studies are associated with
changes in the defect subsystem of ion-implanted layers as these
changes are responsible for the transient enhanced diffusion of
dopant atoms. \cite{Shao_03,Smith_05,Mannino_00,Mirabella_02} For
example, to explain the experimental data, it is assumed
\cite{Gaiduk_03} that elastic stresses arising in the region of
the $\mathrm{SiGe}$ layer buried in silicon and compressively
strained after growth cause the drift of vacancies to this layer.
Thus, accumulation of the vacancies within the layer and their
transformation into nanovoids occur. In Ref. \cite{Mannino_00} the
transient enhanced diffusion of dopant atoms in hyperfine
boron-doped layers created by the molecular-beam epitaxy is
investigated experimentally. It is assumed that the formation of
the clusters of boron atoms with silicon self-interstitials occurs
in these doped regions during thermal treatment. This process and
a number of similar thermal treatments are characterized by the
local absorption of self-interstitials. Thus, investigations
concerning the influence of the local strained regions and local
sinks on the diffusion of nonequilibrium point defects are of
great importance for the next-generation fabrication processes.

\section{Model of stress-mediated diffusion of point defects}

To calculate distributions of the nonequilibrium point defect
concentration in the field of elastic stresses, the stationary
diffusion equation established in Ref. \cite{Velichko_99} can be
used. In the form convenient for numerical solution this equation
can be written as \cite{Velichko_03}

\begin{equation}\label{Eq_Point_Defect_Diffusion}
\begin{array}{c}

\displaystyle {\frac{d }{d x}} \left[ d^{C}(\chi)\displaystyle
{\frac{d \, (a^{d} \tilde{C}) }{d x}} \right]- \displaystyle
{\frac{d (v_{x} \, \tilde {C}) }{d x}} - \displaystyle
\frac{k^{C}(\chi) \, k^{Sp}(x) \, a^{d} \tilde {C} }{l^{2}_{i}}
\\
\\
+ \displaystyle \frac{1+\tilde {g}^{R}}{l^{2}_{i}}=0 \, ,
\\
 \end{array}
\end{equation}

\noindent where

\begin{equation}
\label{Normalized_defect_concentration}
 \tilde {C} = \frac{C^{\times}}{C_{i}^{\times}} \, ,
 \qquad \qquad d^{C}(\chi) = \frac{d(\chi)}{d_{i}} \, ,
\end{equation}

\begin{equation}\label{Carrier_concentration}
    \chi=\frac{C-C_{B}+\sqrt{(C-C_{B})^{2}+4n_{i}^{2}}}{2n_{i}} \, ,
\end{equation}

\begin{equation}\label{Defect_drift_velocity}
  v_{x}= -d^{C}(\chi) \displaystyle \frac{a^{d}}{k_{B}T}
   \displaystyle \frac{d U^{d}}{d x} \, ,
\end{equation}

\begin{equation}\label{Nonequilibrium_generation_rate}
\tilde {g}^{R}= \frac{g^{R}}{g_{i}} \, .
\end{equation}

\noindent Here $C^{\times}$ and $C^{\times}_{i}$ are the
concentration of point defects (vacancies or self-interstitials)
in the neutral charge state and equilibrium value of this
concentration, respectively; $d(\chi)$ and $d_{i}$  are the
effective diffusion coefficient of point defects and intrinsic
diffusivity of these defects, respectively; $C$ is the
concentration of substitutionally dissolved dopant atoms forming a
doped region; $C_{B}$ is the total concentration of dopant atoms
responsible for opposite-type conductivity; $n_{i}$ is the
intrinsic carrier concentration; $k^{C}(\chi)$ is the normalized
to $k_{i}$ concentration dependence of the effective absorption
coefficient for point defects; $k^{Sp}(x)$ is the spatial
distribution of the absorption coefficient; $k_{i}$ and
$\tau_{i}=k_{i}^{-1}$ are the values of absorption coefficient and
average lifetime for point defects in intrinsic semiconductor,
respectively; $l_{i}= \sqrt{d_{i}\tau_{i}}$ is the average
migration length of point defects in intrinsic semiconductor;
$v_{x}$ is the $x$-coordinate projection of the effective drift
velocity of point defects due to elastic stresses; $U^{d}$ is the
potential energy of these defects in the field of elastic
stresses; $g^{R}$ is the rate of the nonequilibrium point defect
generation per unit volume due to the external radiation or
dissolution of extended defects; $g_{i}$ is the rate of thermally
equilibrium generation of point defect in intrinsic semiconductor.
The function $a^{d}=h^{r}/h^{r}_{i}$ takes into account that the
real constants $h^{r}$ for the transition between the charge
states of point defects deviate from their equilibrium values
$h^{r}_{i}$ due to heavy doping effects and elastic stresses.

Eq. (\ref{Eq_Point_Defect_Diffusion}) differs from the defect
diffusion equation used in Ref. \cite{Mirabella_02} because the
drift of mobile species in the field of elastic stresses is
included. On the other hand, from Eq.
(\ref{Eq_Point_Defect_Diffusion}) the distributions of
nonequilibrium point defects can be derived, instead of computing
the equilibrium defect distributions by means of expressions used
in Ref. \cite{Park_95,Chaudhry_97} Eq.
(\ref{Eq_Point_Defect_Diffusion}) is also convenient for numerical
solutions due to the following characteristic features: (i) This
equation describes the diffusion-reaction-drift of nonequilibrium
point defects with different charge states as a whole, although
only the normalized concentration of the neutral defects
$\tilde{C}$ must be obtained to solve the equation. (ii) The
obtained equation takes into account the drift of all charged
species due to the built-in electric field. At the same time,
there is no explicit term describing this drift. (iii) Despite the
fact that the effective coefficients of
Eq.(\ref{Eq_Point_Defect_Diffusion}) represent nonlinear functions
of $\chi$, these functions are smooth and monotone.
\cite{Velichko_99,Velichko_03}

It is also assumed that the mobility of point defects is
significantly greater than the impurity atom mobility, and there
is no change in the processing conditions or the changes are
sufficiently slow. In this case the time derivative of the defect
concentration is close to zero and distributions of point defects
are quasi-stationary with respect to the distributions of dopant
atoms, clusters, extended defects, and also with respect to
changes in the processing conditions.

\section{Numerical solution}

The finite-difference method \cite{Samarskii_01} is used to find a
numerical solution for Eq. (\ref{Eq_Point_Defect_Diffusion}) in
the one-dimensional (1D) domain $[0,x_{B}]$. Following Ref.
\cite{Samarskii_01}, the first term in the left-hand side of Eq.
(\ref{Eq_Point_Defect_Diffusion}) is approximated by a symmetric
difference operator of the second order accuracy on the space
variable $x$. At the same time, the second term is approximated
with the first order accuracy by the asymmetric
forward/backward-difference operator depending on the drift
direction. On the other hand, if the defect flux due to elastic
stresses is comparable to or below the defect flux due to the
concentration gradient, the second term is also approximated by a
symmetric difference operator of the second order accuracy.
Comparison with exact analytical solutions for the particular
cases of point defect diffusion and calculations on the meshes
with different step sizes were carried out to verify the
approximate numerical solution.

\begin{figure}[!ht]
{\includegraphics {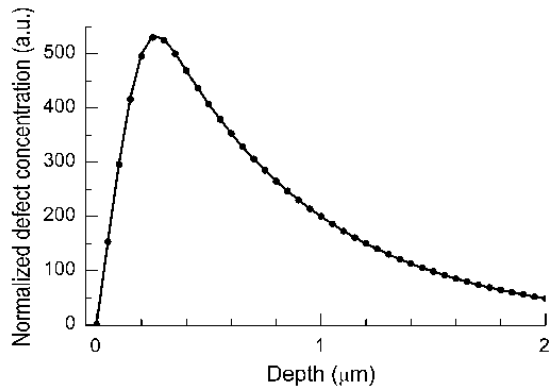}}

\caption{Comparison of the numerical (solid line) and analytical
(dots) solutions for the equation of point defect diffusion.
 \label{fig:test}}.
\end{figure}

For example, in Fig.~\ref{fig:test} the numerical solution for Eq.
(\ref{Eq_Point_Defect_Diffusion}) in case of the constant
diffusivity and constant coefficient of defect absorption is
presented. For comparison with the analytical solution
\cite{Minear_72} Gaussian distribution

\begin{equation}\label{Generation}
\tilde {g}^{R}(x)= \tilde {g}_{m}\exp \left[
-\frac{(x-R_{pd})^{2}}{2\triangle R_{pd}^{ \,2}}\right] \,
\end{equation}

\noindent is used to describe the generation rate profile. Here
$\tilde {g}_{m}$ is a maximum generation rate of point defect
during ion implantation normalized to the thermal generation rate
$g_{i}$; $R_{pd}$ and $\triangle R_{pd}^{2}$ are the position of a
maximum of defect generation distribution and dispersion of this
distribution, respectively. It is supposed that the defect
generation occurs due to implantation of hydrogen ions at energy
20 keV ($R_{pd}$ = 0.198 $\mathrm{\mu m}$ and $\triangle R_{pd}$ =
0.0802 $\mathrm{\mu m}$ are taken from Ref. \cite{Burenkov_85}).
Numerical computations are carried out on the simulation domain
$[0,x_{B}]$, where $x_{B}$ and mesh point number $i_{B}$ are equal
to 4.0 $\mathrm{\mu m}$ and 81, respectively. To obtain a
numerical solution, the Dirichlet boundary conditions are imposed
on $ \tilde{C}$

\begin{equation}\label{Boundary_conditions}
 \tilde{C}(0)=1 \, , \qquad  \qquad \tilde{C}(x_{B})=1 \,.
\end{equation}

The function $\tilde{C}_{th}(x)=1$ is added to the analytical
solution \cite{Minear_72} to satisfy boundary conditions
(\ref{Boundary_conditions}) and take into account the thermal
generation of point defects. As can be seen from
Fig.~\ref{fig:test}, the distribution of nonequilibrium point
defects obtained by numerical computations agrees with the
analytical solution proposed in Ref. \cite{Minear_72}.

\section{Simulation of point defect diffusion}
\begin{figure}[ht]
{\includegraphics {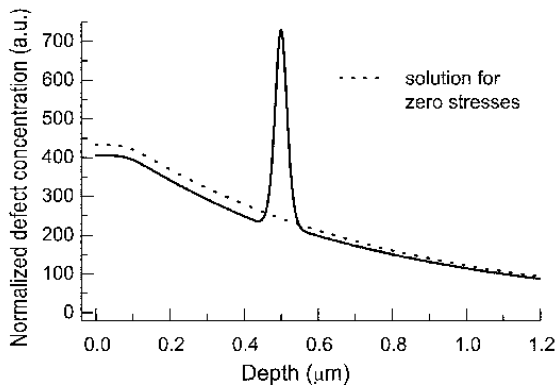}}

\caption{Calculated distribution of the concentration of neutral
point defects passing over the region of local stresses (drift
velocity directed in of the strained layer). A solution for the
same diffusion equation in case of zero stresses is given for
comparison. \label{fig:segregation}}
\end{figure}

In Fig.~\ref{fig:segregation} the calculated distribution of
nonequilibrium point defects in the structure with the local
stress field providing the drift of defects into the strained
region is demonstrated. For comparison, the solution of diffusion
equation (\ref{Eq_Point_Defect_Diffusion}) in case of zero
stresses is also shown by the dotted line.

\begin{figure}[!ht]
{\includegraphics {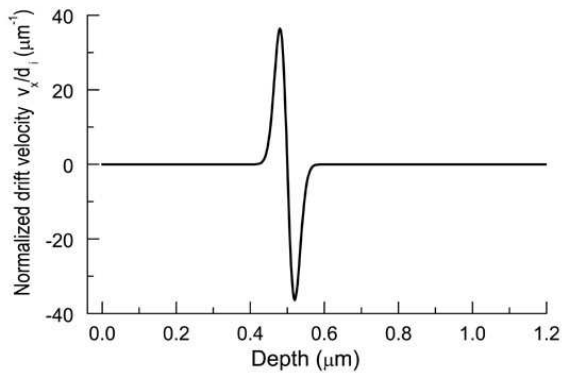}}

\caption{Spatial distribution of the normalized drift velocity of
point defects used in calculations shown in
Fig.~\ref{fig:segregation}.
 \label{fig:stress_segregation}}.
\end{figure}

To meet the experiments with the transient enhanced diffusion, in
all the calculations presented here it is assumed that the
generation of nonequilibrium point defects occurs due to silicon
ion implantation at an energy of 60 keV ($R_{p}$ = 0.081 $\mu$m
and $\triangle R_{p}$ = 0.033 $\mu$m are also from Ref.
\cite{Burenkov_85}), i.e. the generation occurs near the surface
of the semiconductor. Here $R_{p}$ and $\triangle R_{p}$ are the
average projective range of silicon ions and straggling of the
projective range, respectively. To describe the defect diffusion,
the average migration length of point defects $l_{i}$ is chosen
equal to 0.7  $\mu$m to be greater than the thickness of the
strained layer. The reflecting condition for point defects on the
surface and Dirichlet condition in the bulk of semiconductor are
used in all cases under consideration. The distribution of the
normalized drift velocity of point defects in the field of elastic
stresses for the case of defect segregation within the local
strained region (see Fig.~\ref{fig:segregation}) is given in
Fig.~\ref{fig:stress_segregation}. As can be seen from
Fig.~\ref{fig:stress_segregation}, the local field of elastic
stresses takes place between the defect generation region and the
bulk of the semiconductor.

\begin{figure}[ht]
{\includegraphics {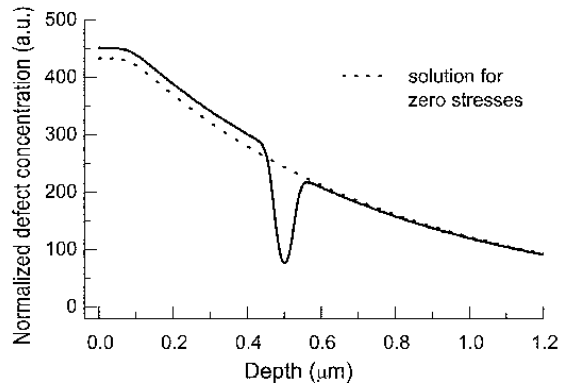}}

\caption{Calculated distribution of the concentration of neutral
point defects passing over the region of local stresses (drift
velocity directed out of the strained layer). A solution for the
same diffusion equation in case of zero stresses is given for
comparison. \label{fig:depletion}}
\end{figure}

\begin{figure}[ht]
{\includegraphics {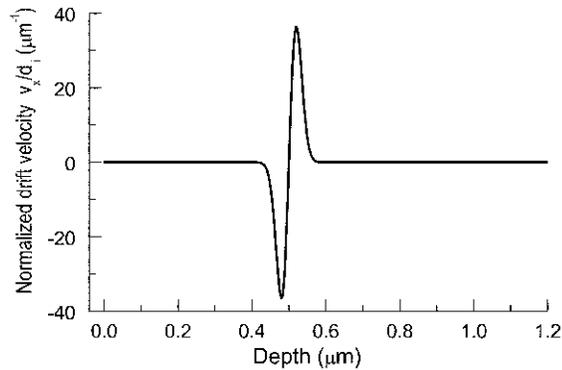}}

\caption{Spatial distribution of the normalized drift velocity of
point defects used in calculations shown in
Fig.~\ref{fig:depletion}.
 \label{fig:stress_depletion}}.
\end{figure}

Fig.~\ref{fig:depletion} presents the calculated distribution of
nonequilibrium point defects in the structure with the local
region of stresses preventing the defect diffusion into the
strained layer. The distribution of the normalized drift velocity
of point defects in the field of elastic stresses used for this
calculation is shown in Fig.~\ref{fig:stress_depletion}. As can be
seen from Fig.~\ref{fig:stress_depletion}, it is supposed that the
local stress field also occupies a position between the defect
generation region and the bulk.

As seen from Figs.~\ref{fig:segregation} and ~\ref{fig:depletion},
the presence of the local stresses results either in enrichment or
depletion of the stress region by point defects. On the other
hand, the distributions of nonequilibrium defects beyond the
region of the local stresses are practically unchanged, regardless
of the stress barrier that should be overcome by the point defects
migrating in the bulk of the semiconductor.

\begin{figure}[!ht]
{\includegraphics {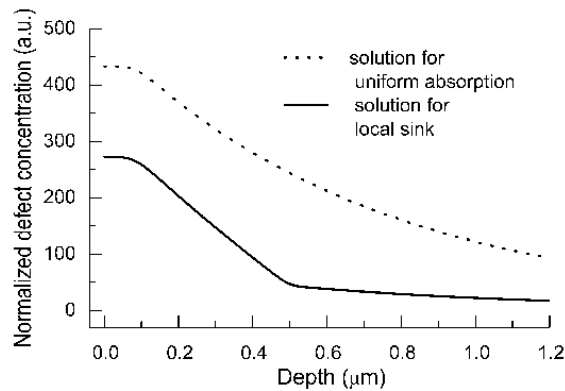}}

\caption{Calculated distribution of the concentration of neutral
point defects passing over the region of local sink. For
comparison a solution of the diffusion equation for the thermally
equilibrium uniform absorption of point defects is also presented.
\label{fig:sink}}
\end{figure}

\begin{figure}[ht]
{\includegraphics {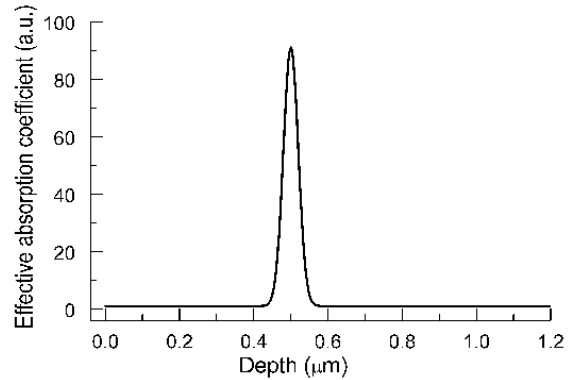}}

\caption{Spatial distribution of the effective coefficient of
point defect absorption used in calculations presented on
Fig.~\ref{fig:sink}.
 \label{fig:absorption}}.
\end{figure}

A qualitatively different situation takes place in the case of
strong local sinks of point defects, for example, in the case of
the silicon structure with an epitaxially grown
$\mathrm{Si_{1-x}C_{x}}$ layer. \cite{Mirabella_02} This can be
seen in Fig.~\ref{fig:sink} showing the point defect concentration
profile calculated for the local sink position at the same place
as the position of local stresses in the previous calculations.
Fig. ~\ref{fig:absorption} demonstrates the spatial distribution
for the effective absorption coefficient of point defects used in
the last calculation.

A significant decrease in defect concentrations in the regions
before and after the sink is easily predictable and agrees with
the experimental data obtained in Ref. \cite{Mirabella_02} It is
interesting to note that the concentrations of nonequilibrium
defects beyond the region of the local stresses can be
significantly changed if the generation/absorption of point
defects takes place in this region. To illustrate, in our previous
work \cite{Velichko_99} it is shown that the generation of silicon
interstitial atoms in the region of the local stresses providing
the drift of point defects from the surface results in
supersaturation of self-interstitials in the bulk of the
semiconductor.

\section{Conclusions}
The influence of local elastic stresses on the formation of point
defect distributions during diffusion of nonequilibrium defects
from the surface into the bulk of the semiconductor has been
investigated by computer simulation. Such local regions of
stresses can be formed in the multilayered heterostructures or
semiconductor structures with heavily doped layers. Thus, point
defects diffusing into the bulk must pass over the region of
stresses if the average defect migration length is greater than
the thickness of the strained layer. Within the strained layer,
the point defect segregation or heavy defect depletion occur if
the defect drift under stresses is directed respectively in or out
of the layer. The numerical calculations show that, in contrast to
the case of local defect sink, the local region of elastic
stresses practically does not change the distribution of defects
beyond this region provided that there is no generation/absorption
of point defects within the strained layers.

\end{document}